\documentclass[aps,pre,superscriptaddress,twocolumn,amsmath,amssymb,showpacs]{revtex4}
\usepackage{graphicx}% Include figure files
\usepackage{dcolumn}% Align table columns on decimal point
\usepackage{bm}% bold math

\begin{document}

\title{Competitive Brownian and L\'evy walkers}
%\shorttitle{Title} %Insert here a short version of the title if it exceeds 80 characters

\author{E.~Heinsalu}
\affiliation{IFISC, Instituto de F\'isica Interdisciplinar y Sistemas Complejos (CSIC-UIB),
  Campus Universitat de les Illes Balears, E-07122 Palma de Mallorca, Spain}
\affiliation{National Institute of Chemical Physics and Biophysics,
  R\"avala 10, Tallinn 15042, Estonia}

\author{E.~Hern\'andez-Garc\'{\i}a}
\affiliation{IFISC, Instituto de F\'isica Interdisciplinar y
Sistemas Complejos (CSIC-UIB),
  Campus Universitat de les Illes Balears, E-07122 Palma de Mallorca, Spain}

\author{C.~L\'opez}
\affiliation{IFISC, Instituto de F\'isica Interdisciplinar y
Sistemas Complejos (CSIC-UIB),
  Campus Universitat de les Illes Balears, E-07122 Palma de Mallorca, Spain}

\date{November 18, 2011}

\begin{abstract}
Population dynamics of individuals undergoing birth and death
and diffusing by short or long ranged twodimensional spatial
excursions (Gaussian jumps or L\'{e}vy flights) is studied.
Competitive interactions are considered in a global case, in
which birth and death rates are influenced by all individuals
in the system, and in a nonlocal but finite-range case in which
interaction affects individuals in a neighborhood (we also
address the noninteracting case). In the global case one single
or few-cluster configurations are achieved with the spatial
distribution of the bugs tied to the type of diffusion. In the
L\'{e}vy case long tails appear for some properties
characterizing the shape and dynamics of clusters. Under
non-local finite-range interactions periodic patterns appear
with periodicity set by the interaction range. This length acts
as a cut-off limiting the influence of the long L\'{e}vy jumps,
so that spatial configurations under the two types of diffusion
become more similar. By dividing initially everyone into
different families and following their descent it is possible
to show that mixing of families and their competition is
greatly influenced by the spatial dynamics.
\end{abstract}


\maketitle

%%%%%%%%%%%%%%%%%%%%%%%%%%%%%%%%%%%%%%%%%%%%%%%%%

\section{Introduction}

%%%%%%%%%%%%%%%%%%%%%%%%%%%%%%%%%%%%%%%%%%%%%%%%%

Birth and death are the most relevant processes in determining
the dynamics of biological populations which in the context of
statistical physics can be modeled using interacting particle
models where particle number is changing in time. As it is
understood by now, birth and death processes are also
responsible for clustering mechanisms in systems where
random-walking individuals undergo reproduction and death. As a
result, aggregation of organisms can occur even in simple
models where birth and death processes are combined with
spatial diffusion. In fact, in the most simple {\it Brownian
bug model}, where particles reproduce and die with the same
probability and undergo Brownian motion
\cite{Zhang1990,Young2001,Felsenstein1975}, clustering of
particles was observed. In this model the clustering is
produced simply by the reproductive correlations (the offspring
is born at the same location of the parent) and by the
irreversibility of the death process. Birth and death models of
moving individuals are the pertinent framework to capture
properties of biological systems such as planktonic populations
\cite{Young2001}, or patterns in amoebae
\cite{Houchmandzadeh2008} and bacteria \cite{Ramos2008}.

Taking into account another central ingredient that is present
in ecological systems, namely, the competition with other
individuals in the neighborhood for resources, the formation of
periodic spatial structures was observed in
Refs.~\cite{Hernandez2004, Hernandez2005, Lopez2004}. In these
{\it nonlocally interacting Brownian bug models} it was assumed
that the reproduction probability depends on the number of
other organisms in the neighborhood. In
Ref.~\cite{Heinsalu2010} {\it nonlocally interacting L\'evy
bugs}, i.e., reproducing and dying organisms that undergo
L\'evy flights, were studied. This type of motion is relevant
to model cell migration~\cite{Dieterich2008}, biological
searching strategies~\cite{Sims2008, Bartumeus2005}, bacteria
dynamics~\cite{levandowsky1997}, or pattern formation of
mussels \cite{deJager2011}. In Ref.~\cite{Heinsalu2010} it was
shown that the formation of a periodic pattern is robust with
respect to the type of spatial motion that the particles
perform. The periodic arrangement of clusters in these
nonlocally interacting bug models is a consequence of the
competitive interaction and has a spatial scale determined by
the interaction range \cite{Hernandez2004}. However, a deeper
analysis of the differences and similarities between the
Brownian and L\'evy cases is still missing. In particular, as
shown in \cite{Hernandez2005,Brigatti}, this analysis can be
very conveniently performed by considering the limit of the
interaction distance reaching the system size (global
interaction), since a unique cluster appears which helps to
understand and characterize the cluster properties, and the
fluctuations of the population size.

In the present paper we report on differences between the systems of
Brownian and L\'evy bugs, in the situations of global and
non-local interactions, as well as in the noninteracting case.
In addition, results on the dependence of population on
diffusion, and mixing of {\it families} of particles are
presented for the finite-range interaction case. The paper is
organized as follows: in Sec.~\ref{sec-model}
we describe the models to be analyzed. In
Sec.~\ref{SecNonInt} the noninteracting bug systems are studied. The
infinite competition range where each particle is competing
with all the others is analyzed in Sec.~\ref{SecGlInt}.
Finally, the nonlocally interacting (i.e. with a finite
interaction range) models are investigated in Sec.~\ref{SecNonLocInt}.

%%%%%%%%%%%%%%%%%%%%%%%%%%%%%%%%%%%%%%%%%%%%%%%%%%

\section{Model and numerical algorithm} \label{sec-model}

%%%%%%%%%%%%%%%%%%%%%%%%%%%%%%%%%%%%%%%%%%%%%%%%%%%

We consider a system consisting initially of $N_0$ point-like
particles, which could be thought as being biological organisms
or bugs, placed randomly in a two-dimensional $L \times L$
square domain with periodic boundary conditions. Except when
explicitly stated, we take $L=1$, so that lengths are measured
in units of system size. Bugs diffuse, reproduce at rate
$r^i_b$, and die at rate $r^i_d$; $i = 1, \ldots, N$, and $N
\equiv N(t)$ is the number of bugs in the system at time $t$.
The numerical algorithm used to evolve the system follows the
one suggested in Ref.~\cite{Birch2006}. The following sequence
of steps is repeated until the final simulation time is
reached:

We first compute the random time $\tau$ after which the next
demographic event (birth or death) will occur. For this we need
to determine the total birth and death rates,
\begin{equation}
B_\mathrm{tot} = \sum _{i=1} ^N r^i_b \, , \qquad D_\mathrm{tot} = \sum _{i=1} ^N r^i_d \, ,
\end{equation}
and compute also the total rate
\begin{equation} \label{R-tot}
R_\mathrm{tot} = B_\mathrm{tot} + D_\mathrm{tot} = \sum _{i=1} ^N (r^i_b + r^i_d) \, .
\end{equation}
For the random times $\tau$ we choose an exponential
probability density with the complementary cumulative distribution
\begin{equation} \label{Ptau-exp}
p(\tau) = \exp(-\tau / \tilde{\tau})
\end{equation}
so that values of $\tau$ could be generated from $\tau =
-\tilde{\tau} \ln(\xi_0)$, where $\xi_0$ is a uniform random
number on $(0, 1)$ \cite{numerical}. The characteristic time or
time-scale parameter $\tilde{\tau}=\langle \tau \rangle$ is
determined by the total rate:
\begin{equation} \label{time-scale}
\tilde{\tau} =  R_\mathrm{tot}^{-1} \, .
\end{equation}

After the random time $\tau$, an individual $i$, chosen among
all the $N(t)$ bugs, either reproduces or disappears. With
probability $B_\mathrm{tot} / R_\mathrm{tot}$ the event is
reproduction and with probability $D_\mathrm{tot} /
R_\mathrm{tot}$ it is death. The probability of choosing a
particular individual $i$ is weighted proportionally to its
contribution to the corresponding total rate. In the case of
reproduction, the new bug is located at the same position
$(x_i, y_i)$ as the parent individual $i$. Finally, all the
bugs perform a jump of random length $\ell$ in a random
direction characterized by an angle uniformly distributed on
$(0, 2 \pi)$ ($\ell$ and the direction of the jump are
independent for each particle). The new present time is  $t +
\tau$, bugs are relabeled with indices $i=1,2,...,N(t+\tau)$,
and the process is repeated.

When bugs undergo normal diffusion (Brownian bugs), a Gaussian
jump-length probability density function is used,
\begin{equation}
\varphi(\ell) = \frac{2}{\tilde{\ell} \sqrt{2\pi}}
\exp \left( -\frac{\ell ^2}{2 {\tilde{\ell}}^2} \right) \, ,\, l\geq 0
\label{GaussianJump}
\end{equation}
with second moment $\langle \ell ^2 \rangle =
{\tilde{\ell}}^2$; $\tilde{\ell}$ is the space-scale parameter.
Since we draw the angle specifying the direction of the jump
from the interval $(0,2\pi)$, we restrict $\ell$ in Eq.
(\ref{GaussianJump}) to have positive sign. The random jump
length $\ell$ can be computed from $\ell = \tilde{\ell} \,
\xi_G $, where $\xi_G$ is sampled from the standard Gaussian
distribution with average $0$ and standard deviation $1$, and
neglecting the sign. Note that the random walk defined in this
way is not exactly the same as the one in which the walker
performs jumps extracted from a two-dimensional Gaussian
distribution, but it also leads to normal diffusion and allows
a more direct comparison with the L\'evy case. The
corresponding diffusion coefficient can be estimated as
\begin{equation} \label{kappa}
\kappa = \langle \ell ^2 \rangle / (2 \langle \tau \rangle) = {\tilde{\ell}}^2 / (2 \tilde{\tau}) \, .
\end{equation}
As we choose to fix the value of $\kappa$, and the demographic
rates, then the space-scale parameter is determined by
\begin{equation}
\tilde{\ell} = \sqrt{2 \kappa \tilde{\tau}} = \sqrt{2 \kappa / R_\mathrm{tot}} \, .
\end{equation}

In order to simulate the system where the bugs undergo
superdiffusive L\'{e}vy flights (L\'evy bugs) one can use a
symmetric L\'evy-type probability density function for the jump
size ($\ell\ge 0$), behaving asymptotically as
\cite{Klages2008,metzler2000}
\begin{equation} \label{levyPDF}
\varphi _\mu (\ell) \approx {\tilde{\ell}}^{\mu}  |\ell|^{-\mu - 1} \, ,
\quad \ell \to \infty \quad (\ell \gg \tilde{\ell})
\end{equation}
with the L\'evy index $0 < \mu < 2$. For all L\'evy-type
probability density functions with $\mu < 2$ the second moment
diverges, $\langle \ell ^2 \rangle = \infty$, leading to the
occurrence of extremely long jumps, and typical trajectories
are self-similar, showing at all scales clusters of shorter
jumps interspersed with long excursions. For $0 < a < \mu < 2$
fractional moments $\langle \ell^a \rangle$ are finite. For the
L\'evy index in the range $1 < \mu < 2$ the value of $\langle
\ell  \rangle$ is finite. The complementary cumulative
distribution corresponding to (\ref{levyPDF}) behaves as
\begin{equation} \label{Plevy}
P_\mu (\ell) \approx {\mu}^{-1} ( \ell  / \tilde{\ell} )^{-\mu} \, ,
\quad \ell \to  \infty \, .
\end{equation}
As a simple form of complementary cumulative distribution
function which behaves asymptotically as (\ref{Plevy}), we use
\begin{equation} \label{Plambda-pareto}
P_\mu (\ell) = (1 + b^{1/\mu} \ell  / \tilde{\ell})^{-\mu} \, ,
\end{equation}
with $b = [\Gamma(1-\mu / 2) \Gamma (\mu / 2)] / \Gamma (\mu)$,
and $\ell \geq 0$. As before, the direction of the jump is
assigned by drawing an random angle on $(0,2\pi)$. The
particular expression for $b$ is chosen for consistency with
previous work \citep{Heinsalu2010}. It gives to the tail of the
jump distribution the same prefactor as for the L\'{e}vy-stable
distribution \citep{Nolan2012}, but any other positive value of
$b$ should lead to the same results reported here. One can
generate a random step-length $\ell$ by inverting
(\ref{Plambda-pareto}):
\begin{equation} \label{Ljump}
\ell = \tilde{\ell} \frac{(\xi _0 ^{-1/\mu} - 1)}{b^{1/\mu}} \, .
\end{equation}
with $\xi _0$ being a uniform random variable on the unit
interval. Now the diffusion coefficient (\ref{kappa}) is
infinite, but one can define a generalized diffusion
coefficient in terms of the scales $\tilde{\ell}$ and
$\tilde{\tau}$ as \cite{metzler2000,Klages2008}
\begin{equation}
\kappa_\mu = {\tilde{\ell}}^\mu / (2 \tilde{\tau}) \, .
\end{equation}
Therefore, in the case of the L\'evy flights, when fixing the value of $\kappa_\mu$, the space-scale
parameter is:
\begin{equation}
\tilde{\ell} = (2 \kappa_\mu \tilde{\tau})^{1/\mu} = (2 \kappa_\mu / R_\mathrm{tot})^{1/\mu} \, .
\end{equation}

As we consider the bugs to be point-like, the spatial dynamics
does not include any interaction between them. The interaction
is instead taken into account through reproduction and death
rates, which we assume to be affected by competitive
interactions.

If the birth and death rates of a bug are influenced by the
number of other bugs within a certain radius $R$, one talks
about a {\it nonlocal interaction} of finite range. In the
present paper we assume that the birth and death rates of the
$i$-th individual depend linearly on the number of neighbors in
the interaction range \cite{Hernandez2004},
\begin{eqnarray}
r^i_b &=& \mathrm{max} \left( 0, r_{b0} - \alpha N_R^i \right) \, , \label{r-birth} \\
r^i_d &=& \mathrm{max} \left( 0, r_{d0} + \beta N_R^i \right) \, . \label{r-death}
\end{eqnarray}
Here $N_R^i$ is the number of bugs which are at a distance
smaller than $R$ from bug $i$, the parameters $r_{b0}$ and
$r_{d0}$ are the zero-neighbor birth and death rates, and the
parameters $\alpha$ and $\beta$ determine how $r^i_b$ and
$r^i_d$ depend on the neighborhood. For positive values of
$\alpha$ and $\beta$, the more neighbors an individual has
within the radius $R$, the smaller is the probability for
reproduction and the larger is the probability that the bug
does not survive, which could arise from competition for
resources. The function $\mathrm{max}()$ enforces the
positivity of the rates. Since we take $R < L/2$ (in fact $R
\ll L$), the periodic boundary conditions are straightforwardly
implemented and bugs are never counted twice.

If the birth and death rates of a bug are instead influenced by
all the other individuals in the system, i.e.,
\begin{eqnarray}
r^i_b &\equiv& r_b = \mathrm{max} \left\{0, r_{b0} - \alpha [N(t) -1 ] \right\} \, , \label{r-birthG} \\
r^i_d &\equiv& r_d = \mathrm{max} \left\{0, r_{d0} + \beta [N(t) -1] \right\} \, , \label{r-deathG}
\end{eqnarray}
then one talks about {\sl global interaction}. This is formally
equivalent to Eqs.~(\ref{r-birth}) and (\ref{r-death}) with $R$
sufficiently large for the interaction domain to include the
whole system, but taking care of counting each bug only once,
so that $N^i_R = N(t) - 1$.

In the case the rates of the demographic events are the same
for all the bugs and assume constant values,
\begin{eqnarray}
r^i_b \equiv r_b = r_{b0} \, ,  \qquad r^i_d \equiv r_d = r_{d0} \, , \label{r-const}
\end{eqnarray}
which is equivalent to $\alpha = \beta = 0$ in
Eqs.~(\ref{r-birth}), (\ref{r-death}), bugs are {\it
noninteracting}.

In the following we discuss the Brownian and L\'evy bug systems
when individuals do not influence each other and when
inter-particle interaction occurs, either global or nonlocal.
Although we formally maintain the parameter $\beta$ in Eqs.~
(\ref{r-death}) and (\ref{r-deathG}), in our numerical examples
we restrict to $\beta=0$.

\begin{figure*}[ht]
\centering
\includegraphics[width=15.0cm]{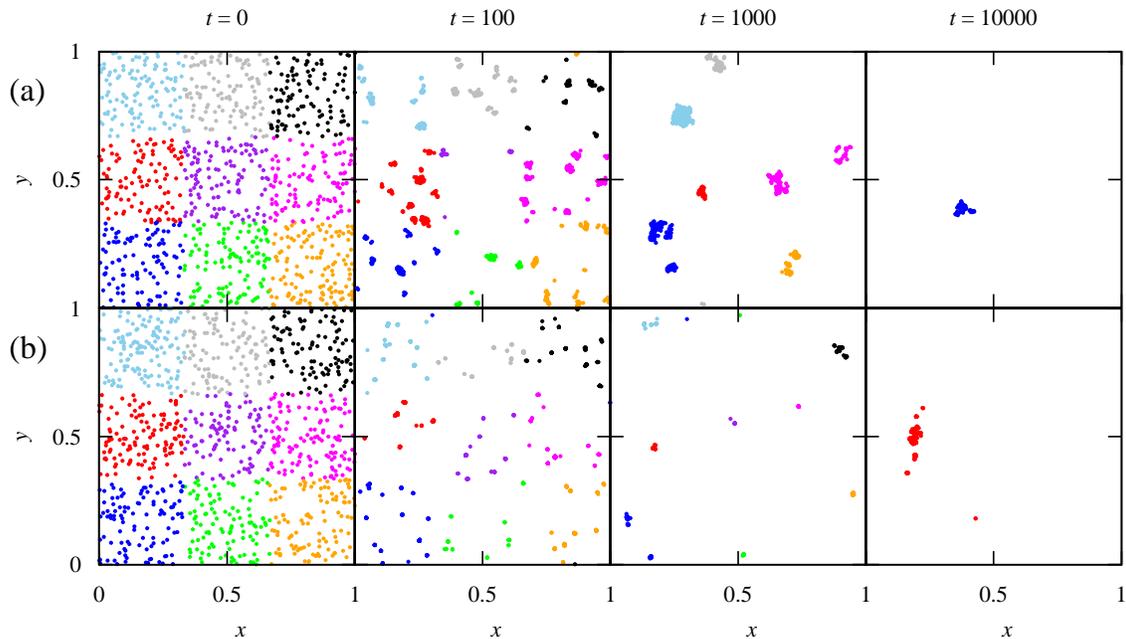}
\caption{(Color) Simple bug models with no interaction; spatial
configuration of bugs at different times $t$: (a) Brownian bugs
with $\kappa = 10^{-6}$ and (b) L\'evy bugs with $\kappa_\mu =
10^{-5}$ and $\mu=1$. Reproduction and death occur with equal
probability, $r_b = r_d = 0.1$, and the initial number of
individuals is $N_0 = 1000$. Bugs are colored with the color
their ancestors bear in the panel at $t=0$.} \label{FigBBxy-6}
\end{figure*}
%

%%%%%%%%%%%%%%%%%%%%%%%%%%%%%%%%%%%%%%%%%%

\section{Simple bug models with no interaction} \label{SecNonInt}

%%%%%%%%%%%%%%%%%%%%%%%%%%%%%%%%%%%%%%%%%%

%%%%%%%%%%%%%%%%%%%%%%%%%%%%%%%%

\subsection{Noninteracting Brownian bugs}

%%%%%%%%%%%%%%%%%%%%%%%%%%%%%%%%

The simple Brownian bug model with no interaction, i.e., when
the birth and death rates of the individuals are given by
Eq.~(\ref{r-const}), has been studied and discussed in various
works \cite{Zhang1990, Young2001}. The ensemble average of the
total population size follows
\begin{equation} \label{number}
\langle N(t) \rangle = N_0 \exp{[\Delta (t - t_0)]} \, ,
\end{equation}
independently of the diffusivity of the bugs; it only depends
on the difference $\Delta = r_b - r_d$. If the birth rate is
larger than the death rate, $\Delta > 0$, the total population
generally explodes exponentially, though there is a finite
probability for extinction that depends on the initial size of
the population and decreases with increasing $\Delta$. If the
death rate is larger than the birth rate, $\Delta < 0$ the
extinction of the population takes place with probability $1$.
If birth and death are equally probable, $\Delta =0$, then the
average over many realizations is $\langle N(t) \rangle = N_0$
and the average lifetime is infinite. However, in single
realizations the fluctuations in the number of individuals are
huge leading to fast extinction in some runs. In fact, there
exists a typical lifetime proportional to $N_0$, defined as the
time for which the fluctuations become as large as the mean
value, beyond which the population is extinct with probability
close to $1$ \cite{Zhang1990}.

As a surprising effect, in the systems where the noninteracting
Brownian bugs undergo death and reproduction with equal
probabilities, spatial clustering of the bugs was observed in
single realizations \cite{Zhang1990,Young2001,Felsenstein1975}.
A typical time evolution of such a system is illustrated in
Fig.~\ref{FigBBxy-6}a. We note that in all figures presenting
the spatial configurations of the bugs, we have divided the
individuals according to their initial position into nine
groups characterized by different colors as in
Fig.~\ref{FigBBxy-6} at time $t = 0$; if reproduction takes
place, the newborn bug assumes the same color as the parent.
From Fig.~\ref{FigBBxy-6}a one can see that many small clusters
form some time after starting from a uniform initial
distribution. The occurrence of the clustering is related to
the fact that in the case of reproduction the new bug is
located at the same position as the parent. Due to the
fluctuations and irreversibility of death the number of
clusters decreases in time, until there will be a single
cluster consisting of individuals coming from a single
ancestor. There are constant, spontaneous, short-lived
break-offs from the main cluster, which are always located near
it. The center of mass of such a cluster undergoes a motion
similar to that of a single bug \cite{Zhang1990} and its linear
width fluctuates with a typical value proportional to
$\sqrt{N_0}$ \cite{Zhang1990}. Furthermore, the larger the
diffusion coefficient $\kappa$, the wider is the cluster
(notice that when simulating the system numerically, if the
diffusivity becomes so large that the jump lengths become
comparable to the system size, one needs to take a larger
simulation box). Finally, due to the fluctuations also the last
cluster disappears.

%%%%%%%%%%%%%%%%%%%%%%%%%%%%%%%%

\subsection{Noninteracting L\'evy bugs}

%%%%%%%%%%%%%%%%%%%%%%%%%%%%%%%%

In the case of noninteracting L\'evy bugs, the number of
individuals still follows Eq.~(\ref{number}), independently of
the L\'evy index $\mu$, and also the cluster formation observed
in the case of Brownian bugs takes place. Now, however, as bugs
can perform long jumps, there are also small clusters
continuously appearing and disappearing far from the main
clusters (Fig.~\ref{FigBBxy-6}b). The smaller the value of
$\mu$ the more anomalous the system, i.e., the larger is the
probability for long jumps and therefore there are more
flash-clusters. When the number of clusters has already
decreased to one, due to the long jumps and fluctuation of the
number of individuals, new clusters that are placed far from
the central cluster can appear in the system also for some time
and often the disappearance of the main cluster takes place
whereas another new central cluster appears somewhere else. As
a result the center of mass undergoes anomalous diffusion as
single bugs do. The value of the L\'evy index $\mu$ influences
also the linear size of the main clusters: the smaller is
$\mu$, the more compact are the clusters, although also more
particle jumps to long distances occur.
 The influence of the value of $\kappa_\mu$ is
similar as in the case of Brownian bugs, i.e., a larger value
of the anomalous diffusion coefficient results in a larger
linear size of the clusters.

\begin{figure*}[t]
\centering
\includegraphics[width=15.0cm]{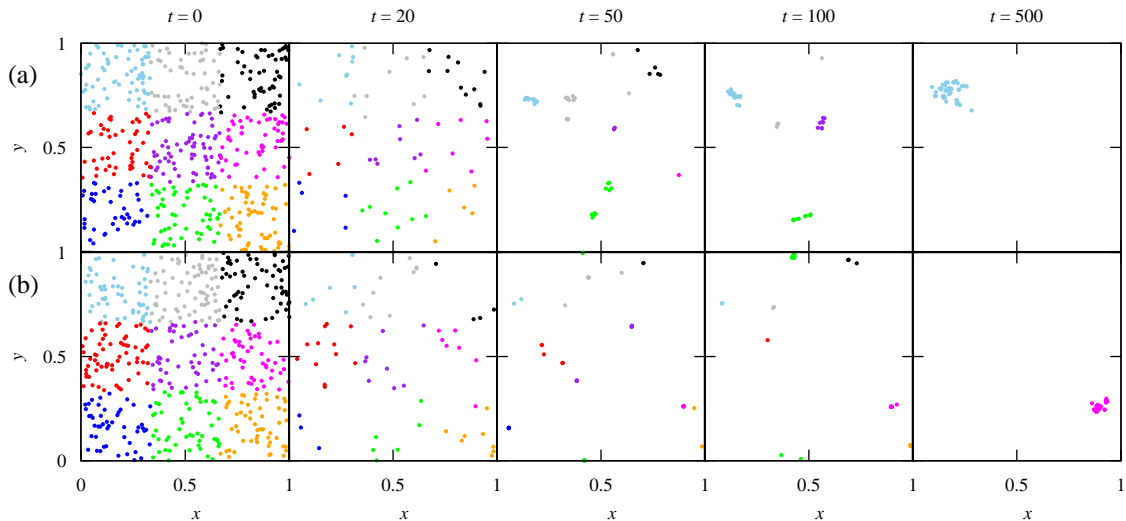}
\caption{(Color) Globally interacting bug models; spatial
configuration of bugs at different times $t$: (a) Brownian bugs
with $\kappa = 10^{-5}$ and (b) L\'evy bugs with $\kappa_\mu =
10^{-4}$ and $\mu=1$. The parameters in the reproduction and
death rates are: $r_{b0} = 1$, $r_{d0} = 0.1$, $\alpha = 0.02$,
$\beta = 0$. Bugs are colored with the color their ancestors
bear in the panel at $t=0$.} \label{Fig-global-XY}
\end{figure*}
%

%%%%%%%%%%%%%%%%%%%%%%%%%%%%%%%%%%%%%%%%%%

\section{Global interaction} \label{SecGlInt}

%%%%%%%%%%%%%%%%%%%%%%%%%%%%%%%%%%%%%%%%%%

%%%%%%%%%%%%%%%%%%%%%%%%%%%%%%%%

\subsection{Formation of a cluster}

%%%%%%%%%%%%%%%%%%%%%%%%%%%%%%%%

Let us now investigate the behavior of the Brownian and L\'evy
bug systems in the case of global interaction, i.e., birth and
death rates of the individuals are given by
Eqs.~(\ref{r-birthG}), (\ref{r-deathG}). The time evolutions of
the globally interacting Brownian and L\'evy bug systems are
illustrated by Fig.~\ref{Fig-global-XY}a and
\ref{Fig-global-XY}b, respectively. In both systems we start
from $N_0 = 500$ bugs uniformly distributed in the simulation
area and choose for the parameters characterizing death and
birth rates the following values: $r_{b0} = 1$, $r_{d0} = 0.1$,
$\alpha = 0.02$, $\beta = 0$. As in the noninteracting case,
the final state of the dynamics is complete extinction, since
there is always a nonvanishing probability for a fluctuation
strong enough to produce that. However, if the number of bugs
in the system is large this happens at very long times
\cite{Doering2005}. Then, there is a long-lived {\sl
quasistable} state for which the average number of individuals
$ \langle N(t) \rangle$ can be estimated from the condition
that death and birth are equally probable, $r_b^i = r_d^i$.
From there,
\begin{equation} \label{global-N}
\langle N(t) \rangle = \frac{\Delta_0 }{\alpha + \beta} +1 \, ,
\end{equation}
where $\Delta _0 = r_{b0} - r_{d0}$. We have restricted to
parameter values so that the $\mathrm{max}$ functions in
Eqs. (\ref{r-birthG}-\ref{r-deathG}) do not operate. Since we have
chosen $N_0
> \langle N(t) \rangle = 46$ in Fig. ~\ref{Fig-global-XY}, death is more probable at small
times and the number of bugs decreases rapidly. Approximately
at time $t = 30$ the number of individuals has reached the
value at which death and birth become in average equally
probable and after this time particle number fluctuates around
that value; parameters of the birth and the death rates can be
chosen so that these fluctuations are weak. At this time small
clusters start to form due to the reproductive pair
correlations. As in the case of noninteracting bugs,
fluctuations and irreversibility of death makes the number of
clusters to decrease in time, although now the fluctuations of
the particle content of the different clusters are correlated
to keep the total number close to the value given by
Eq.~(\ref{global-N}) and the process is faster. Finally a
single cluster consisting of bugs coming from the same ancestor
remains (as stated before, it will also disappear at very long
times due to finite-size fluctuation effects) though there are
also now spontaneous short-lived break-offs from the central
cluster as in the case of noninteracting bugs. The center of
mass of such a cluster is moving in space and its linear size
is a fluctuating quantity. The clustering of the globally
interacting bugs was quantitatively discussed in
Ref.~\cite{Hernandez2005} for the one-dimensional Brownian bug
system.

%%%%%%%%%%%%%%%%%%%%%%%%%%%%%%%%

\subsection{Fluctuations of the number of bugs} \label{SecFlucN}

%%%%%%%%%%%%%%%%%%%%%%%%%%%%%%%%

\begin{figure}[t]
\centering
\includegraphics[width=8.0cm]{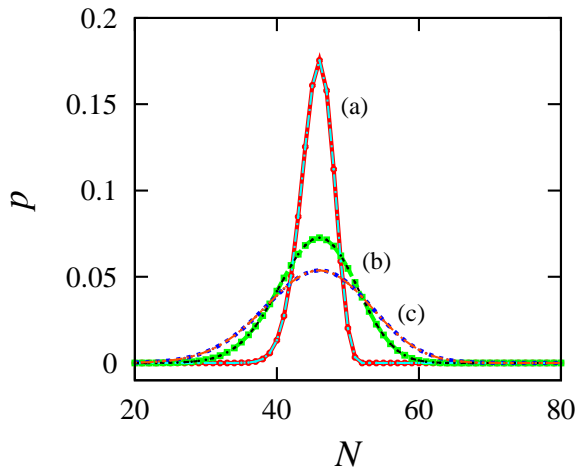}
\caption{(Color online) Probability distribution
of the number of bugs in globally interacting bug systems. The results are numerically obtained
from the time series of the particle number in the very long-lived state before the fluctuations
lead the system to the extinction.
For all the curves $\alpha = 0.02$, $\beta = 0$, and the rate difference is $\Delta _0 = r_{b0} - r_{d0} = 0.9$,
but the rates $r_{b0}$, $r_{d0}$ assume different values:
(a) $r_{b0} = 1$, $r_{d0} = 0.1$; (b) $r_{b0} = 1.5$, $r_{d0} = 0.6$;
(c) $r_{b0} = 2$, $r_{d0} = 1.1$.
The overlapping curves correspond to Brownian and L\'evy bug
systems; the distributions do not depend on the type of diffusion nor on the values of
$\kappa$, $\kappa_\mu$ or $\mu$ in this globally interacting case.}
\label{Fig-Nfluc}
\end{figure}

As indicated by Eq.~(\ref{global-N}), for given values of
$\alpha$ and $\beta$, the average number of individuals in the
system with global interaction depends solely on the difference
$\Delta _0 = r_{b0} - r_{d0}$. It is independent of the
concrete values of $r_{b0}$ and $r_{d0}$, as well as of the
value of $\kappa$ or $\kappa_\mu$ and $\mu$; in fact, it does
not even depend on whether the system consists of Brownian or
L\'evy bugs. Nevertheless, fluctuations of the number of bugs
do indeed depend on the values of $r_{b0}$ and $r_{d0}$, even
if the difference $\Delta _0$, and thus the average number of
bugs, has the same value. To illustrate this, let us calculate
from the simulations time series the probability distribution
of the number of individuals in the globally interacting
Brownian and L\'evy bug systems. As can be seen from
Fig.~\ref{Fig-Nfluc}, for a given value of $\Delta _0$, larger
values of $r_{b0}$ and $r_{d0}$ lead to larger fluctuations.
This is a simple consequence of the Poisson character of the
birth and death events for which fluctuations in each of the
instantaneous rates are proportional to the mean rates. When
the distributions are narrow, they are close to Gaussian. For
larger rates particle number distribution gets broader implying
that there is an enhanced probability that particle number
becomes zero at some moment, after which bugs become extinct
(remember that what is in fact plot in Fig.~\ref{Fig-Nfluc} is
the numerical particle number distribution in the long-lived
metastable state before extinction). For the present case with
$\Delta _0 = 0.9$ and $\alpha = 0.02$, $\beta = 0$, rate values
above the ones shown in Fig.~\ref{Fig-Nfluc} (i.e. $r_{b0}> 2$,
$r_{d0}
> 1.1$) lead to observable extinction after some tenths
of thousands of steps. An ecological implication of this could
be the following: one can think of two colonies of organisms of
the same type, having both the same equilibrium size determined
for example by the size of the living area. Now if in one of
the systems the population has no enemies and the natural death
rate is low, but in the other the death rate is higher due to
the presence of a predator, then the latter system will more
probably go to extinction sooner due to the presence of larger
fluctuations.

%

%%%%%%%%%%%%%%%%%%%%%%%%%%%%%%%%

\subsection{The average cluster shape, cluster width, and center of mass motion} \label{SecAvCl}

%%%%%%%%%%%%%%%%%%%%%%%%%%%%%%%%

\begin{figure}[t]
\centering
\includegraphics[width=8.0cm]{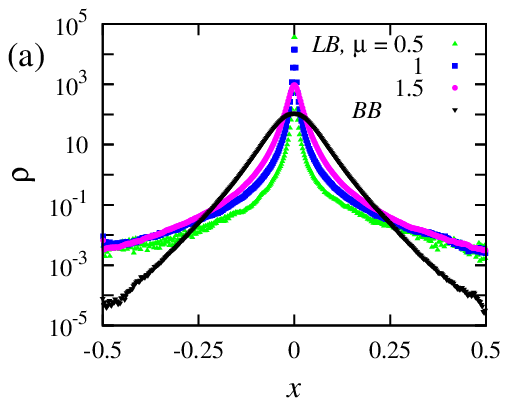} \\
\includegraphics[width=8.0cm]{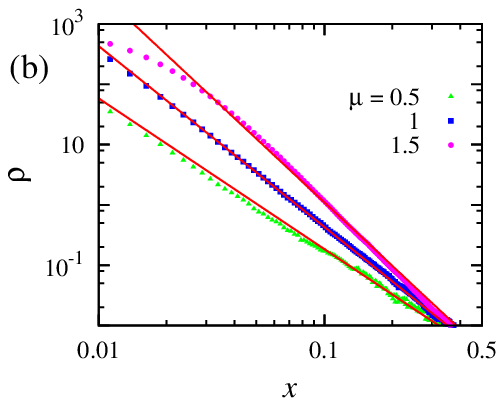}
\caption{(Color online) (a) $\rho(x)$, the cross-section of the two-dimensional
particle density of the average cluster in semi-log scale; comparison
between the Brownian and L\'evy bug systems; $\kappa = 10^{-5}$,
$\kappa _\mu = 10^{-5}$, $r_{b0} = 1$, $r_{d0} = 0.1$, $\alpha = 0.02$, $\beta = 0$.
(b) The tails of $\rho(x)$ in log-log scale in the case of the L\'evy bug systems
for different values of $\mu$. Solid lines correspond to
fitting curves $\propto x^{- (2 + \mu)}$.}
\label{Fig-avcluster}
\end{figure}

\begin{figure}[t]
\centering
\includegraphics[width=8.0cm]{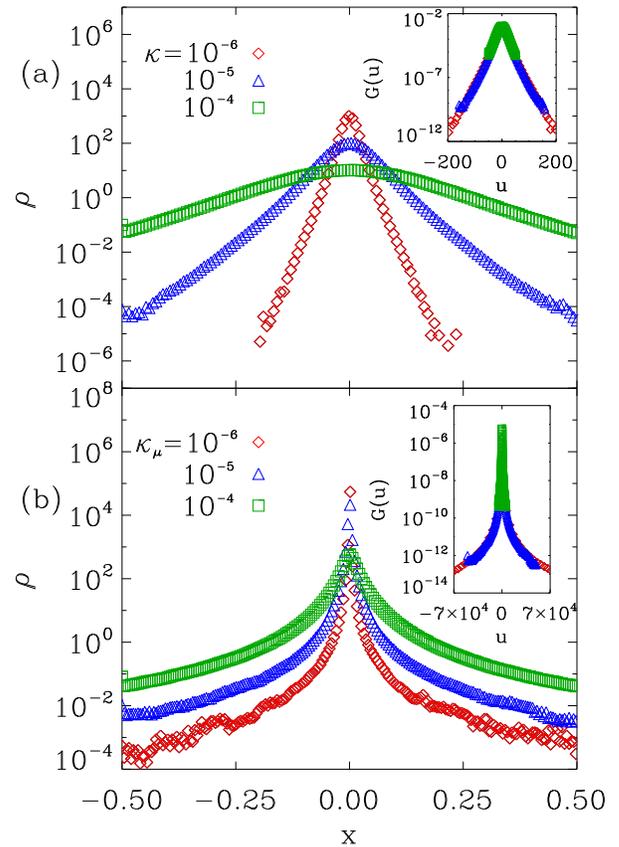}
\caption{(Color online) $\rho(x)$, the cross-section of the two-dimensional
particle density of the average cluster in semi-log scale for different
values of diffusivity: (a) Brownian bugs and (b) L\'evy bugs with $\mu = 1$.
Other parameters are as in Figs.~\ref{Fig-global-XY} and \ref{Fig-avcluster}. The
insets check the correctness of the scaling forms (\ref{clusterscalingG}) (with
$G(u) = \kappa \rho$ and $u = x/\kappa^{1/2}$) and
(\ref{clusterscalingL}) (with $G(u) = \kappa_\mu^{2/\mu}\rho$ and $u = x/\kappa^{1/\mu}$).}
\label{Fig-avcluster-k}
\end{figure}

\begin{figure}[t]
\centering
\includegraphics[width=8.0cm]{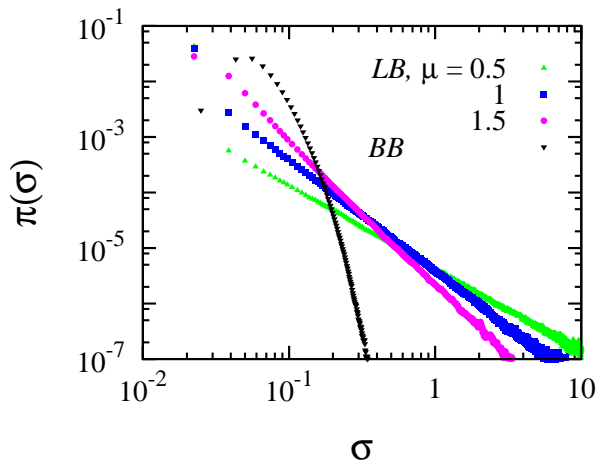}
\caption{(Color online) Probability density $\pi(\sigma)$ of the standard
deviation $\sigma$ of the bug positions with respect to the center of mass of the cluster in the
Brownian and L\'evy bug systems; $\kappa = 10^{-5}$, $\kappa _\mu = 10^{-5}$. The
distribution is obtained averaging over a long time.
The curves corresponding to the L\'evy bug systems are well fitted by
$\propto \sigma^{- (1 + \mu)}$ (not shown).
Other parameters are as in Figs.~\ref{Fig-global-XY}, \ref{Fig-avcluster},
and \ref{Fig-avcluster-k}.}
\label{Fig-MSD}
\end{figure}

\begin{figure}[t]
\centering
\includegraphics[width=8.0cm]{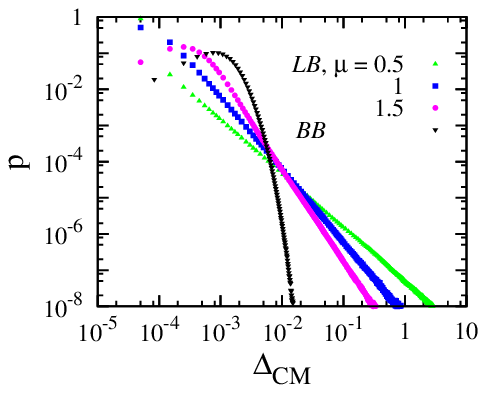}
\caption{(Color online) Probability density $p(\Delta_{CM})$ of the jump lengths
of the center of mass in the Brownian and L\'evy bug systems.
Same parameter values as in Fig.~\ref{Fig-MSD}.
The tails of the curves corresponding to L\'evy bugs are well
fitted by $\propto \Delta_{CM}^{- (1 + \mu)}$ (not shown).}
\label{Fig-JUMP}
\end{figure}

Let us keep in the following $\alpha = 0.02$, $\beta =0$ and
$r_{b0} = 1$, $r_{d0} = 0.1$ [the same parameter values as in
Fig.~\ref{Fig-global-XY} and in Fig.~\ref{Fig-Nfluc} for curve
(a)] and study the behavior of the cluster formed in the case
of a system with global interaction defined by
Eqs.~(\ref{r-birthG})-(\ref{r-deathG}).
 As mentioned, even
after the transition from an uniform distribution of bugs to a
single cluster (and before eventual extinction at large times),
at some moment the system can consist actually of more than one
cluster. In such cases we define that all the individuals in
the system belong to the same cluster, even though in the
L\'evy case the distance between the bugs (subclusters) can be
rather large. In order to avoid the boundary effects as much as
possible, in Figs.~\ref{Fig-avcluster}-\ref{Fig-JUMP} the
linear size of the simulation area was taken as $L = 1000$ and
to have enough statistics simulations were run until $t = 5
\times 10^8$.

Let us start by analyzing the average shape of the cluster. The
average cluster, $\rho(x,y)$, is obtained setting at each time
the origin in the center of mass of the cluster (distances
under the periodic boundary conditions are computed under a
minimum distance convention) and averaging over a long time
(after the transition from uniform distribution to one single
cluster but before long-time extinction). A one-dimensional cut
of it (say across $x$ for $y=0$, i.e., $\rho(x) \equiv \rho(x,
y = 0)$) is shown in Figs.~\ref{Fig-avcluster} and
\ref{Fig-avcluster-k}. For the case of Brownian bugs, the tail
of the average cluster is approximately exponential. A pair
distribution function, which should be related but not
identical to the average cluster discussed here, was
analytically calculated in Ref.~\cite{Birch2006} for a globally
interacting Brownian bug model of our type but in which total
extinction was forbidden. This quantity also displayed a fast
decaying tail. In the case of L\'evy bugs the tail of $\rho(x)$
follows instead a power law, $\rho(x) \sim x^{- (2 + \mu)}$,
see Fig.~\ref{Fig-avcluster}b, arising from the long jumps.
Note that, in the present case of circular symmetry, the
relation $\rho(x,y) dx\, dy = R(r)(2\pi)^{-1} dr \, d\theta$ of
$\rho(x,y)$ with the radial density of the average cluster,
$R(r)$, where $r$ and $\theta$ are the polar coordinates
centered at the cluster center, implies
$\rho(x)=R(r=|x|)(2\pi|x|)^{-1}$, so that the asymptotic
behavior of the radial density is $R(r) \sim r^{- (1 + \mu)}$.
This is the same asymptotic behavior as the individual radial
jumps in (\ref{levyPDF}) and it is also the asymptotic tail of
the probability of displacement from the original position of
nonreproducing bugs moving by L\'evy flights
\cite{metzler2000}. We note also that, for $\kappa = \kappa
_\mu$, the central part of $\rho(x)$ is narrower and higher in
the L\'evy than in the Brownian bug system, and it is narrower
and higher the smaller the value of $\mu$ (see
Fig.~\ref{Fig-avcluster}a). This is a somehow counterintuitive
effect of the L\'{e}vy motion on clusters, already commented in
the noninteracting case: increasing anomalous diffusion
(smaller $\mu$) induces larger jumps and longer tails, but at
small scales it acts as making the cluster more compact.

The influence of the diffusivity is similar in both systems:
the larger is the value of $\kappa$ or $\kappa_\mu$ the more
spread is the average cluster (see Fig.~\ref{Fig-avcluster-k}).
For the Brownian one-dimensional case it was shown in
Ref.~\cite{Hernandez2005} that cluster width is essentially the
distance associated to the Brownian walk during the lifetime of
a bug and their descendants. Thus, the width increases as
$\kappa^{1/2}$. In the L\'{e}vy case, defining the distance
associated to the walk is more subtle, since higher moments of
displacements diverge. But the behavior of lower ones and
dimensional analysis indicate that typical displacements during
a lifetime scale as $\kappa_\mu^{1/\mu}$, and then this should
determine the width of $\rho(x)$ or $R(r)$ (i.e., the spatial
dependence should occur only through the combinations
[$x\kappa_\mu^{-1/\mu}$] or [$r \kappa_\mu^{-1/\mu}$]).
Imposing additionally that the total number of bugs in the
average cluster in this global interaction case does not depend
on particle motion or distribution, and it is thus independent
on the value of $\kappa_\mu$ we have $R(r) =
\kappa_\mu^{-1/\mu} F(r\kappa_\mu^{-1/\mu})$, or
\begin{equation}
\rho(x)=\frac{1}{\kappa_\mu^{2/\mu}}G\left(\frac{x}{\kappa_\mu^{1/\mu}}\right) \ ,
\label{clusterscalingL}
\end{equation}
with $G(u) = F(u)/u$. The analogous scaling form for the average
cluster in the Gaussian diffusion case is
\begin{equation}
\rho(x)=\frac{1}{\kappa}G\left(\frac{x}{\kappa^{1/2}}\right) \ .
\label{clusterscalingG}
\end{equation}
The insets in Fig.~\ref{Fig-avcluster-k} show the validity of
these scaling forms.


In addition to the average cluster shape, which gives
information on the cluster width, it is also interesting to
study the fluctuations of the cluster width in time. We
characterize cluster width at each instant of time by the
standard deviation of the bug positions with respect to the
center of mass of the cluster at that time. Then, a probability
density $\pi(\sigma)$ is constructed from the values of
$\sigma$ at different times. Figure~\ref{Fig-MSD} shows that in
the case of globally interacting Brownian bugs the distribution
of $\sigma$ is short-tailed. In the case of globally
interacting L\'evy bugs, in contrast, the distribution for
$\sigma$ is characterized by tails with a power law decay with
exponent $- (1 + \mu)$. This means that in the latter case the
cluster width can undergo arbitrarily large fluctuations in
time. We note that the tails in $\pi(\sigma)$ decay with the
same exponent as the radial density $R(r)$ of the average
cluster, thus suggesting that the tails of the average cluster
are produced by the large fluctuations in the width of the
instantaneous clusters (which in fact include splitting
events).

The individual motion of bugs drives the behavior of the center
of mass of the system. Figure~\ref{Fig-JUMP} depicts the
probability density $p(\Delta_{CM})$ of the jump lengths
$\Delta_{CM}$ performed by the center of mass each time the bug
motion step is executed in the globally interacting Brownian
and L\'evy bug systems. For Brownian bugs it is short-tailed.
In fact, from the arguments in Ref.~\cite{Hernandez2005}, the
center of mass motion of the cluster for globally interacting
Brownian bugs is characterized by a Brownian process with the
same diffusion coefficient as the individual bugs. In the case
of L\'evy bugs the probability density of the jump lengths of
the center of mass is described by a distribution with a
power-law tail with exponent $-(1 + \mu)$, i.e., the center of
mass of the cluster formed in the case of globally interacting
L\'evy bugs undergoes jumps that follow asymptotically the same
law as the single bugs, Eq.~(\ref{levyPDF}), and as the radial
tails of the average cluster. This reflects the fact commented
previously that, due to the long jumps of the L\'{e}vy bugs,
additional clusters far from the main one appear from time to
time, strongly displacing the center of mass of the system. Due
to the fluctuations it is even possible that the cluster that
used to be the main cluster disappears and a new main cluster
forms somewhere else. As a result the center of mass motion
undergoes the same type of superdiffusion as the individual
bugs of the system.

Extending the arguments for the Brownian bugs
\cite{Hernandez2005} (which were themselves adapted from the
ones in \cite{Zhang1990}) to the L\'{e}vy case one can
heuristically show that the distributions of $\sigma$ and
$\Delta_{CM}$ are related. To this aim one makes the
approximation that the number of bugs in the system is
constant, say $N$, instead of being constant on average. The
center of mass receives a positive contribution from the new
bugs appearing (at location $\vec x_i$) due to the reproduction
between diffusion steps (say at time $t_i$), a negative
contribution from the bugs disappearing during that time (say
from position $\vec x_j$ at time $t_j$), and the direct
contribution from the L\'{e}vy jumps $\vec \ell_k$ of all bugs
present at the diffusion step:
\begin{equation}
\vec\Delta_{CM}=\frac{1}{N} \sum_{i\in B} \vec x_i(t_i) - \frac{1}{N} \sum_{j\in D} \vec x_j(t_j)+
\frac{1}{N} \sum_{k=1}^N \vec\ell_k \ .
\label{CenterOfMassMotion}
\end{equation}
$B$ and $D$ denote the sets of bugs that have been born or
dead, respectively, between diffusion steps. The two first
terms can combined in a single one $\vec S \approx
N^{-1}\sum_{p=1}^n \vec\sigma_p$ by considering that the two
sets have approximately the same number of individuals, $n$.
$\vec\sigma_p = \vec x_i - \vec x_j$ is the displacement
between a pair of these bugs, one just born and the other just
disappeared, sampled inside the same cluster. Then the modulus
of each $\sigma_p$ should be of the order of the cluster width
$\sigma$, which fluctuates in time with probability tails ruled
by an exponent $-(1+\mu)$. This contribution in
Eq.~(\ref{CenterOfMassMotion}) gives the motion of the center
of mass due to the birth and death processes. The contribution
from the individual particle jumps is in the last term in
(\ref{CenterOfMassMotion}), which is a sum of L\'evy jumps of
parameter $\mu$ so that the tails of the probability density
are characterized by a decay with the same exponent $-(1+\mu)$.
These heuristic arguments imply that the modulus $\Delta_{CM}$
will also be distributed with long tails characterized by an
exponent $-(1+\mu)$, as observed.

%%%%%%%%%%%%%%%%%%%%%%%%%%%%%%%%%%%%%%%%%%

\section{Nonlocal interaction} \label{SecNonLocInt}

%%%%%%%%%%%%%%%%%%%%%%%%%%%%%%%%%%%%%%%%%%

%%%%%%%%%%%%%%%%%%%%%%%%%%%%%%%%

\subsection{Formation of a periodic pattern}

%%%%%%%%%%%%%%%%%%%%%%%%%%%%%%%%

%
\begin{figure}[]
\centering
\includegraphics[width=8.0cm]{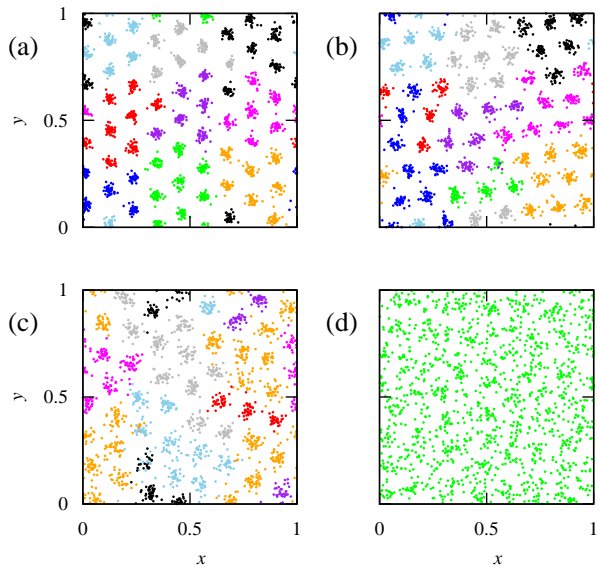}
\caption{(Color) Interacting Brownian bug model
with $R=0.1$, $r_{b0} = 1$, $r_{d0} = 0.1$ and $\alpha = 0.02$, $\beta = 0$.
Spatial configuration of bugs at time $45000$ in systems
with different diffusion coefficients: (a) $\kappa = 10^{-5}$,
(b) $\kappa = 2 \times 10^{-5}$, (c) $\kappa = 4 \times 10^{-5}$,
(d) $\kappa = 10^{-4}$. The initial configuration of bugs is the same as in
Figs.~\ref{FigBBxy-6} and \ref{Fig-global-XY} at time $t = 0$.}
\label{FigIBBxy-4}
\end{figure}

\begin{figure}[]
\centering
\includegraphics[width=8.0cm]{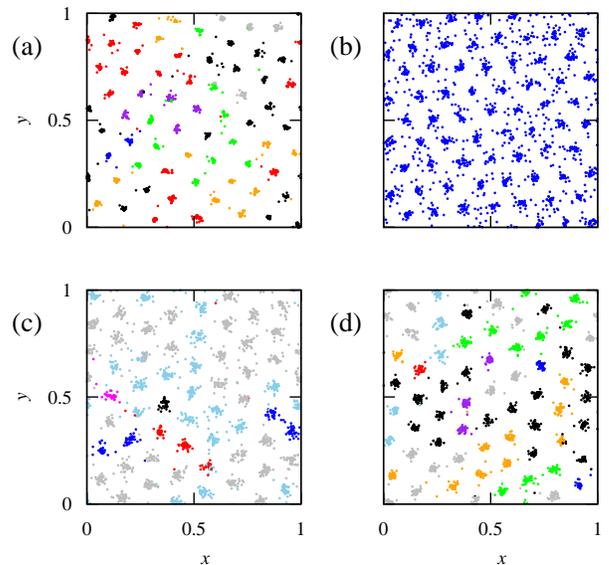}
\caption{(Color) Interacting L\'evy bug model with $R=0.1$, $r_{b0} = 1$,
$r_{d0} = 0.1$ and $\alpha = 0.02$, $\beta = 0$ (same parameters as in Fig.~\ref{FigIBBxy-4} for Brownian bugs).
The spatial configuration of bugs at time $45000$ in systems with
different generalized diffusion coefficients and anomalous exponent:
(a) $\kappa_\mu = 10^{-4}$, $\mu = 1$; (b) $\kappa_\mu = 10^{-3}$, $\mu = 1$;
(c) $\kappa_\mu = 10^{-4}$, $\mu = 1.5$; (d) $\kappa_\mu = 5 \times 10^{-5}$,
$\mu = 1.5$. The initial configuration of bugs is the same as in
Figs.~\ref{FigBBxy-6} and \ref{Fig-global-XY} at time $t = 0$.}
\label{xxx}
\end{figure}

In Refs.~\cite{Hernandez2004, Lopez2004, Heinsalu2010} on the
nonlocally interacting Brownian and L\'evy bugs it was assumed
that the birth and death rates of the $i$-th individual are
given by Eqs.~(\ref{r-birth}), (\ref{r-death}). In the case of
Brownian bugs, for small enough diffusion coefficient and large
enough $\Delta_0$, the occurrence of a periodic pattern
consisting of clusters that are arranged in a hexagonal lattice
was observed (see Fig.~\ref{FigIBBxy-4}a-c)
\cite{Hernandez2004, Lopez2004}. For large values of the
diffusion coefficient such periodic pattern is replaced by a
more homogeneous distribution of bugs (Fig.~\ref{FigIBBxy-4}d).
In the case of L\'evy bugs, since the diffusion coefficient
(\ref{kappa}) is infinite, one could expect that the spatial
distribution will not reveal a periodic pattern; however, as
shown in Ref.~\cite{Heinsalu2010}, for proper parameters
periodic cluster arrangements do indeed occur (see
Fig.~\ref{xxx}). The reason for the divergence of the diffusion
coefficient in the L\'evy case is in the statistical weight of
large jumps. These large jumps have some influence on the
characteristics of the pattern formed, but the relevant
structure is ruled mainly by the interactions between
individuals. In the L\'evy bug system however, at variance with
the Brownian case, even at small values of $\kappa_\mu$ there
are many solitary bugs appearing for short time periods in the
space between the periodically arranged clusters due to the
large jumps \cite{Heinsalu2010}, c.f. Figs.~\ref{FigIBBxy-4}a
and \ref{xxx}a. However, the periodicity of the pattern is of
the order of $R$ (the interaction range) in both systems, being
only slightly influenced by $\kappa$ or $\kappa _\mu$ and
$\mu$, as demonstrated in Refs.~\cite{Hernandez2004,
Heinsalu2010} through a mean-field theory calculation.

In Ref.~\cite{Heinsalu2010} also the two-dimensional particle
density of the average cluster, obtained by setting the origin
at the center of mass of each cluster forming the periodic
pattern and averaging over all the clusters in the simulation
area and over time, was studied. In both, Brownian and L\'evy
bug systems the central part of the average cluster, where most
of the individuals are concentrated, was well fitted by a
Gaussian function, but the way the particle density decreases
when moving away from the center of mass of the cluster is
rather different. In the Brownian case a Gaussian decay
provides a good approximation, whereas in the L\'evy case it is
close to exponential. The comparison with the systems with
global interaction, discussed in Sec.~\ref{SecAvCl}, reveals
therefore that the interaction range $R$ turns the exponential
decay into Gaussian and the power law decay into exponential.

For a given value of diffusion coefficient, there exists a
critical value of $\Delta _0$ below which the system gets
extinct, independently of $\alpha$ \cite{Hernandez2004}. Above
this critical value, for every $\alpha$ the increase of $\Delta
_0$ results in the increase of the average number of bugs, but
the pattern formation is not much influenced. The latter is,
however, true solely if $\Delta _0$ increases through the
increase of $r_{b0}$ and the death rate is low. Namely, as in
the case of global interaction discussed in
Sec.~\ref{SecFlucN}, an increase of the death rate, though
accompanied by a compensating increase of birth rate, leads to
larger fluctuations in the particle number. In numerical
simulations we have observed that the larger are the
fluctuations in the number of bugs, the more difficult is the
formation of the periodic pattern, and finally the individuals
do not arrange in the periodic pattern but in random clusters
(see also Ref.~\cite{Birch2006}). This effect may in fact make
difficult to observe the periodic clustering phenomenon in real
competitive biological systems.

\begin{figure}[]
\centering
\includegraphics[width=8.0cm]{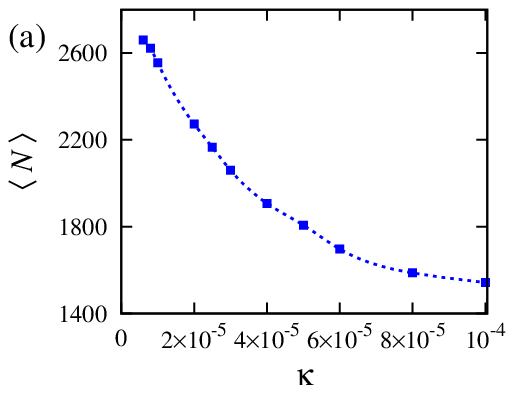}
\includegraphics[width=8.0cm]{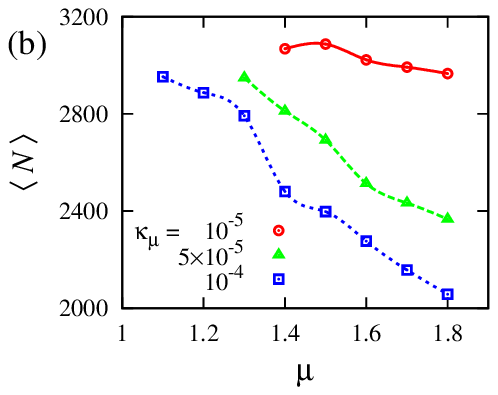}
\caption{(Color online) a): Average number of bugs
versus diffusion coefficient in
the system with Gaussian jumps.
b): Average number of bugs versus anomalous exponent $\mu$
in the system with L\'evy
jumps for various values of the
anomalous diffusion coefficient. Other parameters as in Figs. \ref{FigIBBxy-4} and \ref{xxx}. }
\label{FigNavG}
\end{figure}

In the following we keep for the parameters in the birth and
death rate the same values as in the case of global
interaction, i.e., $r_{b0} = 1$, $r_{d0} = 0.1$, $\alpha =
0.02$, $\beta =0$. For these parameter values the number of
bugs fluctuates only weakly around the mean value. Differently
from the case of global interaction, now the average number of
bugs in the system is influenced not only by the birth and
death rates, but also by the diffusion, i.e., in the case of
Brownian bugs by $\kappa$ and in the case of L\'evy bugs by
$\kappa_\mu$ and $\mu$, see Fig.~\ref{FigNavG}. The smaller is
$\kappa$, $\kappa_\mu$ or $\mu$, the larger the particle
number. At the same time Figs.~\ref{FigIBBxy-4} and \ref{xxx}
indicate that by decreasing $\kappa$ or $\kappa _\mu$ the
linear width of the clusters becomes smaller, the particle
density in the clusters higher, and the density between the
clusters lower (c.f. Sec.~\ref{SecAvCl} and see also
Ref.~\cite{Heinsalu2010}). Somehow counterintuitively, the
effect of decreasing $\mu$ seems to have the same effects, as
commented above for the noninteracting and global cases.
Furthermore, the value of $\kappa$ or $\kappa_\mu$ and $\mu$
seems to weakly influence the number of clusters in the system:
In Figs. and \ref{FigIBBxy-4} and \ref{xxx} smaller values lead
to larger number of clusters. This observation is not explained
by the linear instability analysis of
\cite{Hernandez2004,Heinsalu2010}.

%

%%%%%%%%%%%%%%%%%%%%%%%%%%%%%%%%

\subsection{Mixing of different families}

%%%%%%%%%%%%%%%%%%%%%%%%%%%%%%%%

It is interesting to study the evolution of the system also
regarding the disappearance or survival of the different
groups, by dividing initially the bugs into different families
and following their descent. In the case of nonlocally
interacting Brownian bugs, a very low diffusion coefficient
leads to the situation in which after cluster formation the
inter-cluster travel is very rare because the individuals are
not capable to make the jumps from one cluster to another one,
and it is also very unlikely to arrive to the next cluster
doing a multistep random walk because death is very probable
between the clusters. Therefore, in the case of very low
diffusion different families would remain inside their initial
clusters. If one assumes that initially each individual
represents a different family, then only inter-cluster
competition occurs and the final number of families is equal to
the number of clusters. If instead initially individuals are
assigned to families according to large areas of initial
positions (larger than typical cluster size as done in Figs.
\ref{FigBBxy-6} and \ref{Fig-global-XY} at time $t = 0$), there
is no family competition internal to the clusters, most
families survive and the clusters coming from different
families occupy approximately the territory of the ancestors
even after a long time, as can be seen from
Fig.~\ref{FigIBBxy-4}a. In that case, the travel of a cluster
to a new territory away from the other clusters of the same
family can take place due to the diffusion of the cluster as a
whole during the clusters arrangement into the periodic
pattern. For larger values of $\kappa$ the inter-cluster travel
is possible which leads to the conquering of new territories,
i.e., bugs can be found in a region where their ancestors were
not from, Fig.~\ref{FigIBBxy-4}b. The effect is larger for
larger $\kappa$ and leads to the disappearance of some
families, as can be seen from Fig.~\ref{FigIBBxy-4}c. Finally,
for increased diffusion, intra-cluster competition will force
all surviving bugs to be from a single family (in fact, from a
single ancestor); which family (ancestor) wins is a random
event. The process is faster for larger diffusion. Increasing
the diffusivity further even the periodic pattern disappears,
Fig.~\ref{FigIBBxy-4}d.

Figure~\ref{xxx} illustrates the family mixing for nonlocally
interacting L\'evy bugs. In this case the inter-cluster
traveling is supported by the long jumps. Differently from the
case of Brownian bugs, now the individuals can reach not only
the next neighboring cluster but also clusters far away.
Consequently, a cluster originally consisting of bugs coming
from one ancestor can after some time turn into a cluster
consisting of bugs coming from different families placed
initially far away. Thus, intra-cluster bug competition becomes
soon competition between families, and even if the diffusivity
of the bugs is very low, at the end the L\'evy bug system would
consist of individuals coming from one or just a few ancestors.
As in the case of Brownian bugs, the process of the
disappearance of families is faster the greater is the
generalized diffusion coefficient.

Besides the diffusion of a cluster as a whole during the
formation of the periodic pattern and the conquering of new
territories through the migration to and survival  in another
cluster, the mixing of clusters from different families can
take place also due to the appearance of a new cluster  if in
the periodic pattern there is a dislocation. In the case of
Brownian bugs the new cluster is formed through the splitting
of an old cluster. In the case of L\'evy bugs, instead, the new
cluster can appear also far from the original territory.

%%%%%%%%%%%%%%%%%%%%%%%%%%%%%%%%%%%%%%%%%%

\section{Conclusions and outlook} \label{conclusion}

%%%%%%%%%%%%%%%%%%%%%%%%%%%%%%%%%%%%%%%%%%

We have presented some detailed properties of interacting
particle systems in which the individuals are Brownian or
L\'evy random walkers which interact in a competitive manner.
We have seen strong differences between the globally and the
finite-range nonlocally interacting systems. In the systems
with global interaction the spatial distribution of the bugs
becomes tied to the type of diffusion, Brownian or L\'{e}vy.
Typical configurations consist of a single or a few clusters
for both types of motion. For the L\'{e}vy bug systems long
tails appear in the mean cluster shape and in probability
distributions of cluster width and of jumps of the center of
mass. For Brownian bug systems these quantities appear to be
much shorter ranged. This is qualitatively also the situation
in the noninteracting case, although then the effects of the
particle-number fluctuations are much stronger. Under non-local
finite-range interactions the situation is rather different.
First, single cluster configurations are generally replaced by
periodic patterns with periodicity set by the interaction range
$R$. Motion of individual clusters is severely restricted by
the presence of the neighboring clusters. In addition, the
natural spatial cut-off introduced by the interaction range $R$
seems to limit the influence of the long L\'{e}vy jumps, so
that measures of spatial cluster shape do not generally exhibit
power laws, making spatial configurations under both types of
diffusion more similar. Mixing of families and their
competition is nevertheless greatly influenced by the type of
motion. This suggests that it would be interesting to consider
the influence of different types of diffusion into competitive
genetic mixing processes \cite{Sayama2002}.

Obtaining analytic understanding in this type of interacting
systems is difficult, but at least the nature of the
instability leading to pattern formation and its relevant
spatial scale have been clarified by using partial
integro-differential equation descriptions of the mean field
type \cite{Hernandez2004,Heinsalu2010}, which are useful in
broader contexts \cite{Fuentes2003,Clerc2005,Maruvka2006}.
However, from previous work in the Gaussian case
\cite{Lopez2004,HernandezPA2005,Hernandez2005}, it is known
that quantities such as cluster width and structure or
transition thresholds strongly depend on particle-number
fluctuations. Thus, obtaining additional results from
differential equation approaches would need the inclusion of
effective multiplicative noise terms \cite{Ramos2008} or focus
on statistical quantities such as pair correlation functions
\cite{Young2001,Birch2006,Houchmandzadeh2009}.

%%%%%%%%%%%%%%%%%%%%%%%%%%%%%%%%%%%%%%%%%%

\acknowledgments
This work has been supported by the targeted
financing project SF0690030s09, Estonian Science Foundation
through grant no. 7466, by the Balearic Government (E.H.),
and by Spanish MICINN and FEDER through project FISICOS
(FIS2007-60327).

%%%%%%%%%%%%%%%%%%%%%%%%%%%%%%%%%%%%%%%%%%

%%%%%%%%%%%%%%%%%%%%%%%%%%%%%%%%%%%%%%%%%%

%\bibliography{Refs-Heinsalu}

\begin{thebibliography}{27}
\expandafter\ifx\csname natexlab\endcsname\relax\def\natexlab#1{#1}\fi
\expandafter\ifx\csname bibnamefont\endcsname\relax
  \def\bibnamefont#1{#1}\fi
\expandafter\ifx\csname bibfnamefont\endcsname\relax
  \def\bibfnamefont#1{#1}\fi
\expandafter\ifx\csname citenamefont\endcsname\relax
  \def\citenamefont#1{#1}\fi
\expandafter\ifx\csname url\endcsname\relax
  \def\url#1{\texttt{#1}}\fi
\expandafter\ifx\csname urlprefix\endcsname\relax\def\urlprefix{URL }\fi
\providecommand{\bibinfo}[2]{#2}
\providecommand{\eprint}[2][]{\url{#2}}

\bibitem[{\citenamefont{Zhang et~al.}(1990)\citenamefont{Zhang, Serva, and
  Polikarpov}}]{Zhang1990}
\bibinfo{author}{\bibfnamefont{Y.-C.} \bibnamefont{Zhang}},
  \bibinfo{author}{\bibfnamefont{M.}~\bibnamefont{Serva}}, \bibnamefont{and}
  \bibinfo{author}{\bibfnamefont{M.}~\bibnamefont{Polikarpov}},
  \bibinfo{journal}{J.~Stat. Phys.} \textbf{\bibinfo{volume}{58}},
  \bibinfo{pages}{849} (\bibinfo{year}{1990}).

\bibitem[{\citenamefont{Young et~al.}(2001)\citenamefont{Young, Roberts, and
  Stuhne}}]{Young2001}
\bibinfo{author}{\bibfnamefont{W.~R.} \bibnamefont{Young}},
  \bibinfo{author}{\bibfnamefont{A.~J.} \bibnamefont{Roberts}},
  \bibnamefont{and} \bibinfo{author}{\bibfnamefont{G.}~\bibnamefont{Stuhne}},
  \bibinfo{journal}{Nature} \textbf{\bibinfo{volume}{412}},
  \bibinfo{pages}{328} (\bibinfo{year}{2001}).

\bibitem[{\citenamefont{Felsenstein}(1975)}]{Felsenstein1975}
\bibinfo{author}{\bibfnamefont{J.}~\bibnamefont{Felsenstein}},
  \bibinfo{journal}{The American Naturalist} \textbf{\bibinfo{volume}{109}},
  \bibinfo{pages}{359} (\bibinfo{year}{1975}).

\bibitem[{\citenamefont{Houchmandzadeh}(2008)}]{Houchmandzadeh2008}
\bibinfo{author}{\bibfnamefont{B.}~\bibnamefont{Houchmandzadeh}},
  \bibinfo{journal}{Phys. Rev. Lett.} \textbf{\bibinfo{volume}{101}},
  \bibinfo{pages}{078103} (\bibinfo{year}{2008}).

\bibitem[{\citenamefont{Ramos et~al.}(2008)\citenamefont{Ramos, L\'opez,
  Hern\'andez-Garc\'ia, and Mu{\~n}oz}}]{Ramos2008}
\bibinfo{author}{\bibfnamefont{F.}~\bibnamefont{Ramos}},
  \bibinfo{author}{\bibfnamefont{C.}~\bibnamefont{L\'opez}},
  \bibinfo{author}{\bibfnamefont{E.}~\bibnamefont{Hern\'andez-Garc\'ia}},
  \bibnamefont{and} \bibinfo{author}{\bibfnamefont{M.~A.}
  \bibnamefont{Mu{\~n}oz}}, \bibinfo{journal}{Phys. Rev.~E}
  \textbf{\bibinfo{volume}{77}}, \bibinfo{pages}{021102}
  (\bibinfo{year}{2008}).

\bibitem[{\citenamefont{Hern\'andez-Garc\'{\i}a and
  L\'opez}(2004)}]{Hernandez2004}
\bibinfo{author}{\bibfnamefont{E.}~\bibnamefont{Hern\'andez-Garc\'{\i}a}}
  \bibnamefont{and} \bibinfo{author}{\bibfnamefont{C.}~\bibnamefont{L\'opez}},
  \bibinfo{journal}{Phys. Rev.~E} \textbf{\bibinfo{volume}{70}},
  \bibinfo{pages}{016216} (\bibinfo{year}{2004}).

\bibitem[{\citenamefont{Hern\'andez-Garcia and L\'opez}(2005)}]{Hernandez2005}
\bibinfo{author}{\bibfnamefont{E.}~\bibnamefont{Hern\'andez-Garcia}}
  \bibnamefont{and} \bibinfo{author}{\bibfnamefont{C.}~\bibnamefont{L\'opez}},
  \bibinfo{journal}{J.~Phys.: Condens. Matter} \textbf{\bibinfo{volume}{17}},
  \bibinfo{pages}{S4263} (\bibinfo{year}{2005}).

\bibitem[{\citenamefont{L\'opez and Hern\'andez-Garcia}(2004)}]{Lopez2004}
\bibinfo{author}{\bibfnamefont{C.}~\bibnamefont{L\'opez}} \bibnamefont{and}
  \bibinfo{author}{\bibfnamefont{E.}~\bibnamefont{Hern\'andez-Garcia}},
  \bibinfo{journal}{Physica D} \textbf{\bibinfo{volume}{199}},
  \bibinfo{pages}{223} (\bibinfo{year}{2004}).

\bibitem[{\citenamefont{Heinsalu et~al.}(2010)\citenamefont{Heinsalu,
  Hern\'andez-Garcia, and L\'opez}}]{Heinsalu2010}
\bibinfo{author}{\bibfnamefont{E.}~\bibnamefont{Heinsalu}},
  \bibinfo{author}{\bibfnamefont{E.}~\bibnamefont{Hern\'andez-Garcia}},
  \bibnamefont{and} \bibinfo{author}{\bibfnamefont{C.}~\bibnamefont{L\'opez}},
  \bibinfo{journal}{Europhys. Lett.} \textbf{\bibinfo{volume}{92}},
  \bibinfo{pages}{40011} (\bibinfo{year}{2010}), \bibinfo{note}{{E}rratum: {\bf
  95}, 69902 (2011)}.

\bibitem[{\citenamefont{Dieterich et~al.}(2008)\citenamefont{Dieterich, Klages,
  Preuss, and Schwab}}]{Dieterich2008}
\bibinfo{author}{\bibfnamefont{P.}~\bibnamefont{Dieterich}},
  \bibinfo{author}{\bibfnamefont{R.}~\bibnamefont{Klages}},
  \bibinfo{author}{\bibfnamefont{R.}~\bibnamefont{Preuss}}, \bibnamefont{and}
  \bibinfo{author}{\bibfnamefont{A.}~\bibnamefont{Schwab}},
  \bibinfo{journal}{Proc. Natl. Acad. Sci. USA} \textbf{\bibinfo{volume}{105}},
  \bibinfo{pages}{459} (\bibinfo{year}{2008}).

\bibitem[{\citenamefont{Sims et~al.}(2008)\citenamefont{Sims, Southall,
  Humphries, Hays, Bradshaw, Pitchford, James, Ahmed, Brierley, Hindell
  et~al.}}]{Sims2008}
\bibinfo{author}{\bibfnamefont{D.}~\bibnamefont{Sims}},
  \bibinfo{author}{\bibfnamefont{E.}~\bibnamefont{Southall}},
  \bibinfo{author}{\bibfnamefont{N.}~\bibnamefont{Humphries}},
  \bibinfo{author}{\bibfnamefont{G.}~\bibnamefont{Hays}},
  \bibinfo{author}{\bibfnamefont{C.}~\bibnamefont{Bradshaw}},
  \bibinfo{author}{\bibfnamefont{J.}~\bibnamefont{Pitchford}},
  \bibinfo{author}{\bibfnamefont{A.}~\bibnamefont{James}},
  \bibinfo{author}{\bibfnamefont{M.}~\bibnamefont{Ahmed}},
  \bibinfo{author}{\bibfnamefont{A.}~\bibnamefont{Brierley}},
  \bibinfo{author}{\bibfnamefont{M.}~\bibnamefont{Hindell}},
  \bibnamefont{et~al.}, \bibinfo{journal}{Nature}
  \textbf{\bibinfo{volume}{451}}, \bibinfo{pages}{1098} (\bibinfo{year}{2008}).

\bibitem[{\citenamefont{Bartumeus et~al.}(2005)\citenamefont{Bartumeus, Luz,
  Viswanathan, and Catal\'an}}]{Bartumeus2005}
\bibinfo{author}{\bibfnamefont{F.}~\bibnamefont{Bartumeus}},
  \bibinfo{author}{\bibfnamefont{M.~G. E.~D.} \bibnamefont{Luz}},
  \bibinfo{author}{\bibfnamefont{G.~M.} \bibnamefont{Viswanathan}},
  \bibnamefont{and}
  \bibinfo{author}{\bibfnamefont{J.}~\bibnamefont{Catal\'an}},
  \bibinfo{journal}{Ecology} \textbf{\bibinfo{volume}{86}},
  \bibinfo{pages}{3078} (\bibinfo{year}{2005}).

\bibitem[{\citenamefont{Levandowsky et~al.}(1997)\citenamefont{Levandowsky,
  White, and Schuster}}]{levandowsky1997}
\bibinfo{author}{\bibfnamefont{M.}~\bibnamefont{Levandowsky}},
  \bibinfo{author}{\bibfnamefont{B.~S.} \bibnamefont{White}}, \bibnamefont{and}
  \bibinfo{author}{\bibfnamefont{F.~L.} \bibnamefont{Schuster}},
  \bibinfo{journal}{Acta Protozool.} \textbf{\bibinfo{volume}{36}},
  \bibinfo{pages}{237} (\bibinfo{year}{1997}).

\bibitem[{\citenamefont{de~Jager et~al.}(2011)\citenamefont{de~Jager, Weissing,
  Herman, Nolet, and van~de Koppel}}]{deJager2011}
\bibinfo{author}{\bibfnamefont{M.}~\bibnamefont{de~Jager}},
  \bibinfo{author}{\bibfnamefont{F.~J.} \bibnamefont{Weissing}},
  \bibinfo{author}{\bibfnamefont{P.}~\bibnamefont{Herman}},
  \bibinfo{author}{\bibfnamefont{B.~A.} \bibnamefont{Nolet}}, \bibnamefont{and}
  \bibinfo{author}{\bibfnamefont{J.}~\bibnamefont{van~de Koppel}},
  \bibinfo{journal}{Science} \textbf{\bibinfo{volume}{332}},
  \bibinfo{pages}{1551} (\bibinfo{year}{2011}).

\bibitem[{\citenamefont{Brigatti et~al.}(2008)\citenamefont{Brigatti,
  Schwammle, and Neto}}]{Brigatti}
\bibinfo{author}{\bibfnamefont{E.}~\bibnamefont{Brigatti}},
  \bibinfo{author}{\bibfnamefont{V.}~\bibnamefont{Schwammle}},
  \bibnamefont{and} \bibinfo{author}{\bibfnamefont{M.~A.} \bibnamefont{Neto}},
  \bibinfo{journal}{Phys. Rev.~E} \textbf{\bibinfo{volume}{77}},
  \bibinfo{pages}{021914} (\bibinfo{year}{2008}).

\bibitem[{\citenamefont{Birch and Young}(2006)}]{Birch2006}
\bibinfo{author}{\bibfnamefont{D.~A.} \bibnamefont{Birch}} \bibnamefont{and}
  \bibinfo{author}{\bibfnamefont{W.~R.} \bibnamefont{Young}},
  \bibinfo{journal}{Theoretical Population Biology}
  \textbf{\bibinfo{volume}{70}}, \bibinfo{pages}{26} (\bibinfo{year}{2006}).

\bibitem[{\citenamefont{Press et~al.}(1992)\citenamefont{Press, Flannery,
  Teulolsky, and Vetterling}}]{numerical}
\bibinfo{author}{\bibfnamefont{W.~H.} \bibnamefont{Press}},
  \bibinfo{author}{\bibfnamefont{B.~P.} \bibnamefont{Flannery}},
  \bibinfo{author}{\bibfnamefont{S.~A.} \bibnamefont{Teulolsky}},
  \bibnamefont{and}
  \bibinfo{author}{\bibfnamefont{W.}~\bibnamefont{Vetterling}},
  \emph{\bibinfo{title}{Numerical {R}ecipes in {C}: {T}he {A}rt of {S}cientific
  {C}omputing}} (\bibinfo{publisher}{Cambridge University Press},
  \bibinfo{address}{Cambridge, 2 edition}, \bibinfo{year}{1992}).

\bibitem[{\citenamefont{Klages et~al.}(2008)\citenamefont{Klages, Radons, and
  Sokolov}}]{Klages2008}
\bibinfo{author}{\bibfnamefont{R.}~\bibnamefont{Klages}},
  \bibinfo{author}{\bibfnamefont{G.}~\bibnamefont{Radons}}, \bibnamefont{and}
  \bibinfo{author}{\bibfnamefont{I.~M.} \bibnamefont{Sokolov}},
  \emph{\bibinfo{title}{{A}nomalous {T}ransport: {F}oundations and
  {A}pplications}} (\bibinfo{publisher}{Wiley-VCH}, \bibinfo{year}{2008}).

\bibitem[{\citenamefont{Metzler and Klafter}(2000)}]{metzler2000}
\bibinfo{author}{\bibfnamefont{R.}~\bibnamefont{Metzler}} \bibnamefont{and}
  \bibinfo{author}{\bibfnamefont{J.}~\bibnamefont{Klafter}},
  \bibinfo{journal}{Phys. Rep.} \textbf{\bibinfo{volume}{339}},
  \bibinfo{pages}{1} (\bibinfo{year}{2000}).

\bibitem[{\citenamefont{Nolan}(2012)}]{Nolan2012}
\bibinfo{author}{\bibfnamefont{J.~P.} \bibnamefont{Nolan}},
  \emph{\bibinfo{title}{Stable Distributions - Models for Heavy Tailed Data}}
  (\bibinfo{publisher}{Birkhauser}, \bibinfo{address}{Boston},
  \bibinfo{year}{2012}), \bibinfo{note}{to appear. Chapter 1 online at
  http://academic2.american.edu/$\sim$jpnolan}.

\bibitem[{\citenamefont{Doering et~al.}(2005)\citenamefont{Doering, Sargsyan,
  and Sander}}]{Doering2005}
\bibinfo{author}{\bibfnamefont{C.~R.} \bibnamefont{Doering}},
  \bibinfo{author}{\bibfnamefont{K.~V.} \bibnamefont{Sargsyan}},
  \bibnamefont{and} \bibinfo{author}{\bibfnamefont{L.~M.}
  \bibnamefont{Sander}}, \bibinfo{journal}{Multiscale Modeling \& Simulation}
  \textbf{\bibinfo{volume}{3}}, \bibinfo{pages}{283} (\bibinfo{year}{2005}).

\bibitem[{\citenamefont{Sayama et~al.}(2002)\citenamefont{Sayama, {de Aguiar},
  Bar-Yam, and Baranger}}]{Sayama2002}
\bibinfo{author}{\bibfnamefont{H.}~\bibnamefont{Sayama}},
  \bibinfo{author}{\bibfnamefont{M.~A.~M.} \bibnamefont{{de Aguiar}}},
  \bibinfo{author}{\bibfnamefont{Y.}~\bibnamefont{Bar-Yam}}, \bibnamefont{and}
  \bibinfo{author}{\bibfnamefont{M.}~\bibnamefont{Baranger}},
  \bibinfo{journal}{Phys. Rev. E} \textbf{\bibinfo{volume}{65}},
  \bibinfo{pages}{051919} (\bibinfo{year}{2002}).

\bibitem[{\citenamefont{Fuentes et~al.}(2003)\citenamefont{Fuentes, Kuperman,
  and Kenkre}}]{Fuentes2003}
\bibinfo{author}{\bibfnamefont{M.~A.} \bibnamefont{Fuentes}},
  \bibinfo{author}{\bibfnamefont{M.~N.} \bibnamefont{Kuperman}},
  \bibnamefont{and} \bibinfo{author}{\bibfnamefont{V.~M.}
  \bibnamefont{Kenkre}}, \bibinfo{journal}{Phys. Rev. Lett.}
  \textbf{\bibinfo{volume}{91}}, \bibinfo{pages}{158104}
  (\bibinfo{year}{2003}).

\bibitem[{\citenamefont{Clerc et~al.}(2005)\citenamefont{Clerc, Escaff, and
  Kenkre}}]{Clerc2005}
\bibinfo{author}{\bibfnamefont{M.~G.} \bibnamefont{Clerc}},
  \bibinfo{author}{\bibfnamefont{D.}~\bibnamefont{Escaff}}, \bibnamefont{and}
  \bibinfo{author}{\bibfnamefont{V.~M.} \bibnamefont{Kenkre}},
  \bibinfo{journal}{Phys. Rev. E} \textbf{\bibinfo{volume}{72}},
  \bibinfo{pages}{056217} (\bibinfo{year}{2005}).

\bibitem[{\citenamefont{Maruvka and Shnerb}(2006)}]{Maruvka2006}
\bibinfo{author}{\bibfnamefont{Y.~E.} \bibnamefont{Maruvka}} \bibnamefont{and}
  \bibinfo{author}{\bibfnamefont{N.~M.} \bibnamefont{Shnerb}},
  \bibinfo{journal}{Phys. Rev. E} \textbf{\bibinfo{volume}{73}},
  \bibinfo{pages}{011903} (\bibinfo{year}{2006}).

\bibitem[{\citenamefont{Hern\'andez-Garc\'{\i}a and
  L\'opez}(2005)}]{HernandezPA2005}
\bibinfo{author}{\bibfnamefont{E.}~\bibnamefont{Hern\'andez-Garc\'{\i}a}}
  \bibnamefont{and} \bibinfo{author}{\bibfnamefont{C.}~\bibnamefont{L\'opez}},
  \bibinfo{journal}{Physica A} \textbf{\bibinfo{volume}{356}},
  \bibinfo{pages}{95 } (\bibinfo{year}{2005}).

\bibitem[{\citenamefont{Houchmandzadeh}(2009)}]{Houchmandzadeh2009}
\bibinfo{author}{\bibfnamefont{B.}~\bibnamefont{Houchmandzadeh}},
  \bibinfo{journal}{Phys. Rev. E} \textbf{\bibinfo{volume}{80}},
  \bibinfo{pages}{051920} (\bibinfo{year}{2009}).

\end{thebibliography}
%\end{document}

%%\bibliographystyle{elsart-num} revtex4 will call the correct apsrev4-1.bst file.

\end{document}